\documentclass[amsmath,superscriptaddress,showpacs,aps,prl,twocolumn]{revtex4-1}
\usepackage[colorlinks,linkcolor=blue,anchorcolor=blue,citecolor=blue,urlcolor=blue]{hyperref}
\usepackage{amsmath}
\usepackage{amssymb}
\usepackage{graphicx}
\usepackage{color}
\usepackage{bm}

\begin{document}
\title{Itinerant topological magnons in Haldane Hubbard model with a nearly-flat electron band}
\author{Zhao-Long Gu}
\affiliation{National Laboratory of Solid State Microstructures and Department of Physics, Nanjing University, Nanjing 210093, China}
\author{Zhao-Yang Dong}
\affiliation{Department of Applied Physics, Nanjing University of Science and Technology, Nanjing 210094, China.}
\affiliation{National Laboratory of Solid State Microstructures and Department of Physics, Nanjing University, Nanjing 210093, China}
\author{Shun-Li Yu}
\affiliation{National Laboratory of Solid State Microstructures and Department of Physics, Nanjing University, Nanjing 210093, China}
\affiliation{Collaborative Innovation Center of Advanced Microstructures, Nanjing University, Nanjing 210093, China}
\author{Jian-Xin Li}
\email[]{jxli@nju.edu.cn}
\affiliation{National Laboratory of Solid State Microstructures and Department of Physics, Nanjing University, Nanjing 210093, China}
\affiliation{Collaborative Innovation Center of Advanced Microstructures, Nanjing University, Nanjing 210093, China}
\date{\today}

\begin{abstract}

\par We elaborate the first theoretical realization of two dimensional itinerant topological magnons, based on the quarter filled Haldane-Hubbard model with a nearly-flat electron band. By using the exact diagonalization method with a projection onto this band, we obtain the spin wave excitations over the itinerant ferromagnetic ground state. In the flatband limit, the excitation exhibits similar dispersion to the free electron band with Dirac magnons. The nonflatness of the electron band opens a topological gap at Dirac points and leads to an acoustic magnon band with a nonzero Chern number. We further show that tuning the sublattice Hubbard interactions or the next-nearest-neighbor hopping can induce a topological transition characterized by the gap closing and reopening, and the existence of the in-gap magnons on magnetic domain walls. We find an exact set of bases for magnons in the flatband limit constructed from sublattice particle-hole vectors and derive an effective model to explore the origin of the topological magnon which is attributed to the ``mass inversion mechanism''.

\end{abstract}

\maketitle

\par Band structures with nontrivial topology reside at the center of a substantial number of topological phenomena in condensed matter physics \cite{Hasan_RMP2010,Qi_RMP2011,Bansil_RMP2016}. They exhibit fascinating physics \cite{Chiu_RMP2016,Zeng_B2019}, and cannot be distinguished from trivial ones by local order parameters, but are characterized by nonzero bulk topological indices and gapless edge states on open boundaries \cite{Laughlin_PRB1981,Halperin_PRB1982,Kane_PRL2005a,Bernevig_S2006,Yu_PRL2011}. It was proposed in a pioneering work by Haldane \cite{Haldane_PRL1988} that a spinless fermionic model on a honeycomb lattice bears energy bands with nonzero Chern numbers \cite{Thouless_PRL1982,Simon_PRL1983} in the absence of external magnetic fields. In this model, the nonzero Chern number arises from the mechanism that the mass terms of the two chirality-opposite Dirac fermions in the Brillouin zone (BZ) have different signs. Microscopically, this so-called ``mass inversion mechanism'' is realized by the introduction to the model of a complex next-nearest-neighbor hopping that breaks the time-reversal symmetry locally. Similar mechanism also applies to other fermionic systems with nontrivial topological bands belonging to different symmetry classes \cite{Kane_PRL2005}.

\par Recently, great interests are drawn to the study of correlated topological states where the presence of strong Coulomb interactions between electrons leads to richer physics \cite{Hohenadler_JPCM2013,Wen_RMP2017,Rachel_RPP2018}, especially when the electron bands are nearly flat so that interaction effects are highly enhanced \cite{Neupert_PS2015}. In fractionally-filled strongly-correlated nearly-flat topological bands, some intriguing topological phases of matter, such as fractional Chern insulators \cite{Tang_PRL2011,Wang_PRB2011,Sun_PRL2011,Wang_PRL2011,Neupert_PRL2011,Sheng_NC2011,Regnault_PRX2011} and fractional topological insulators \cite{Neupert_PRB2011}, can be stabilized even at high temperatures. Besides these emergent states in the framework of the charge degree of freedom which constitutes the main focus of previous studies, the states related to the spin degree of freedom are also of fundamental importance. Actually, in most works that consider spinful electron models with nearly-flat bands, the itinerant ferromagnetism \cite{Tasaki_PRL1992,Mielke_PLA1993,Mielke_CMP1993} of electron spins in the ground state plays as the prerequisite of possible fractional Chern insulators \cite{Tang_PRL2011,Wang_PRB2011,Sun_PRL2011} or integer quantum Hall insulators \cite{Neupert_PRL2012}. However, the researches \cite{Doretto_PRB2015,Su_PRB2019} on the spin excitations over the ferromagnetic ground state, which can uncover new physics of strongly-correlated nearly-flat topological bands, are still far from sufficient.

\par In fact, collective spin excitations over magnetically ordered ground state also exhibit band structures, therefore, exotic magnon excitations with nontrivial band topology are expected to emerge in quantum magnets. Indeed, recently, in a number of local spin materials mostly with Dzyaloshinskii-Moriya (DM) interactions \cite{Dzyaloshinsky_JPCS1958,Moriya_PR1960}, topological magnons have been verified to exist both theoretically \cite{Zhang_PRB2013,Owerre_JPCM2016,Li_NC2016,Mook_PRL2016,Laurell_PRL2017} and experimentally \cite{Onose_S2010,Chisnell_PRL2015,Yao_NP2018,Bao_NC2018}. In such models the magnonic excitations are well understood as free bosons in the framework of linear spin wave theory (LSWT), with the DM term acting as the vector potential for the propagation of magnons. However, in the case of itinerant magnets, LSWT fails due to the lack of exact one local electron spin per physical site. The absence of an analytical effective model describing the spin wave excitations makes the investigations on itinerant topological magnons quite arduous \cite{Su_PRB2018}.

\par In this letter, we elaborate the first theoretical realization of two dimensional itinerant topological magnons which are different from previous ones existing in local spin models and derive an effective model for the collective spin excitations to explain the underlying mechanism leading to the nontrivial magnonic topology. The microscopic model is the spinful Haldane-Hubbard model. The phase diagram of this model at half-filling with $\phi=\pi/2$ has been extensively studied \cite{He_PRB2011a,Maciejko_PRB2013,Zheng_PRB2015,Hickey_PRL2016,Wu_PRB2016,Vanhala_PRL2016,Giuliani_PRB2016,Gu_NJP2019}. Here, we consider the quarter-filled case with a nearly-flat lower electron band in which a finite Hubbard interaction can lead to an itinerant ferromagnetic ground state. By using the numerical exact diagonalization method with a projection onto this electron band, we obtain the spin wave excitations over the ferromagnetic ground state. The spectra host Dirac magnons in the flatband limit. Remarkably, the nonflatness of the electron band can induce a topological gap of the Dirac magnons, leading to an acoustic magnon band with a nonzero Chern number. We also show that tuning the imbalance of the AB sublattice Hubbard interactions or the next-nearest-neighbor hopping can close and reopen the magnon gap, accompanied by a Chern number changing. Consistent with the bulk-edge correspondence, there always exist in-gap magnon states on magnetic domain walls.
Furthermore, we find an exact set of bases for the itinerant spin waves in the flat band limit constructed from sublattice particle-hole vectors, and reduce greatly the projected Hamiltonian into an effective model represented by a $2\times2$ matrix. This simplification of the model in its analytic form makes it possible to understand the origin of the topological magnons which is attributed to the ``mass inversion mechanism''.

\begin{figure}
\centering
\includegraphics[width=0.48\textwidth]{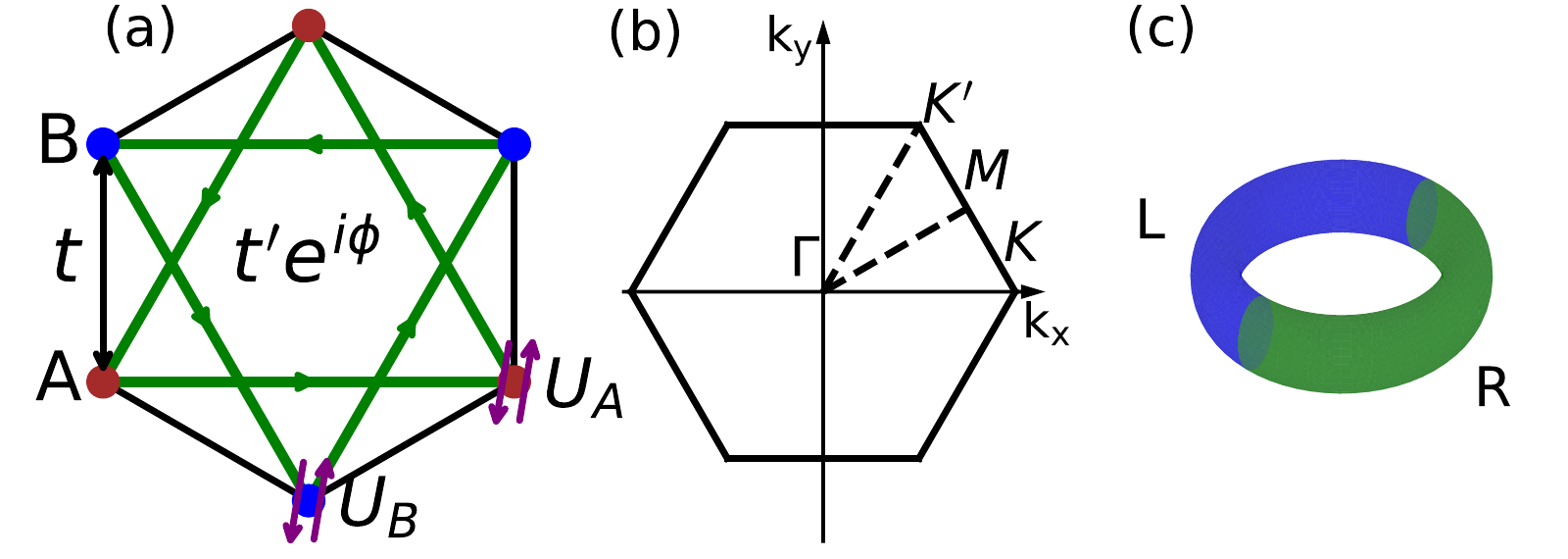}
\caption{(color online). (a) Illustration of the Haldane-Hubbard model on the honeycomb lattice. A and B denote the two inequivalent sites within a unitcell. The real nearest-neighbor hopping $t$, the complex next-nearest-neighbor hopping $t^\prime e^{i\phi}$, and the Hubbard interactions $U_A$ and $U_B$ are also shown. (b) The Brillouin zone. (c) An illustration of the domain wall geometry.}
\label{lattice}
\end{figure}

\par The Haldane Hubbard model as shown schematically in Fig. \ref{lattice}(a) is written by,
\begin{equation}\label{H0}
\hat{H}=t\sum_{\langle ij\rangle\sigma}c^{\dagger}_{i\sigma}c_{j\sigma}+t'\sum_{\langle\langle ij\rangle\rangle\sigma}e^{\phi_{ij}}c^{\dagger}_{i\sigma}c_{j\sigma}+
\sum_i U_i n_{i\uparrow}n_{i\downarrow}.
\end{equation}
Where the first two terms represent the spinful version of the Haldane model \cite{Haldane_PRL1988}, and the third term the Hubbard interaction. $\langle ij\rangle$ and $\langle\langle ij\rangle\rangle$ denote the nearest-neighbor (NN) and next-nearest-neighbor (NNN) bonds, respectively. $\phi_{ij}=\pm\phi$ is the phase of NNN hopping with the sign given by the solid green arrows in Fig. \ref{lattice}(a). $U_i=U_A$ ($U_B$) when $i$ site is the A(B) sublattice. Others are in standard notation.

\par The free part of the Hamiltonian in the momentum space can be written as $\hat{H}_0=\sum_{\mathbf{k}\sigma}\psi^{\dagger}_{\mathbf{k}\sigma}H_0(\mathbf{k})\psi_{\mathbf{k}\sigma}$. Here, $\psi^{\dagger}_{\mathbf{k}\sigma}=(c^{\dagger}_{A\mathbf{k}\sigma},c^{\dagger}_{B\mathbf{k}\sigma})$, and $H_0(\mathbf{k})=\sum_{\alpha=0,x,y,z}h_\alpha(\mathbf{k})\tau^\alpha$, with $\tau^\alpha$ ($\alpha=0,x,y,z$) the identity and Pauli matrices in the sublattice space. When $\phi=0$, then $h_z(\mathbf{k})=0$, the electronic energy band has two Dirac points with opposite chiralities at $K/K^\prime$ points [see Fig. \ref{lattice}(b)]. When $t^\prime\ne0$ and $\phi\ne0$, a gap opens at the Dirac points and $h_z(\mathbf{k})\tau^z$ is the mass term. Because $h_z(K)$ and $h_z(K^\prime)$ have different signs \cite{Haldane_PRL1988}, the Chern numbers of the two massive Dirac points do not cancel each other but add up to a nonzero integer \cite{Rachel_RPP2018}. This is the so-called ``mass inversion mechanism'' for generating a nonzero Chern number (Details see the supplement material).

\par With a proper tuning of the amplitude $t^\prime$ and phase $\phi$ of the NNN hopping, the lower electron band of $\hat{H}_0$ can be quite flat. The flatness ratio $\Delta/W$, which is defined as the ratio of the gap $\Delta$ between the two bands to the bandwidth $W$ of the lower one, takes its maximum ($\sim7$) when $\cos\phi=t/4t^\prime=3\sqrt{3/43}$ ($t^\prime/t\simeq0.3155, \phi\simeq0.656$) \cite{Neupert_PRL2011}. It is well-known that the ground state of such a system is the ferromagnetic state $|\text{FM}\rangle\equiv\prod_{\mathbf{k}\in\text{BZ}}d^\dagger_{\mathbf{k}\uparrow}|0\rangle$ when the Hubbard interaction exceeds a critical value \cite{Tasaki_PRL1994,Su_PRB2019}. Here, $|0\rangle$ is the electron vacuum, and $d^\dagger_{\mathbf{k}\uparrow}$ creates a spin-up electron with a momentum $\mathbf{k}$ in the lower electron band. The parameter space is restricted to the region where $\Delta$ is larger than both $U_A$ and $U_B$, so that the physics is dominated by the degrees of freedom of this lower band and the whole Hamiltonian $H$ can be projected onto it \cite{Neupert_PRL2011,Regnault_PRX2011,Neupert_PRB2011,Neupert_PRL2012,Su_PRB2019}. Thus, a basis of the spin-1 excitations with a center-of-mass momentum $\mathbf{q}$ can be written as $|\mathbf{k}_i\rangle_\mathbf{q}=d^\dagger_{\mathbf{k}_i-\mathbf{q}\downarrow}d_{\mathbf{k}_i\uparrow}|\text{FM}\rangle$. Then, the matrix element of the projected Hamiltonian on this set of bases is
\begin{equation}\label{PHP}
_\mathbf{q}\langle\mathbf{k}_j|P^\dagger \hat{H}P|\mathbf{k}_i\rangle_\mathbf{q}=\left[M_{i}^1(\mathbf{q})+M_{i}^2(\mathbf{q})\right]\delta_{\mathbf{k}_j,\mathbf{k}_i}-M_{ji}^3(\mathbf{q})
\end{equation}
where, $P$ is the projector onto the lower band, and
\begin{eqnarray}
M_{i}^1(\mathbf{q}) &=& \varepsilon_d(\mathbf{k}_i-\mathbf{q})-\varepsilon_d(\mathbf{k}_i), \label{M1}\\
M_{i}^2(\mathbf{q}) &=& \frac{1}{N}\sum_{a=A,B}U_a\sum_{\mathbf{p}}\left|\mu_{a\mathbf{p}\uparrow}\right|^2\left|\mu_{a\mathbf{k}_{i}-\mathbf{q}\downarrow}\right|^2, \label{M2}\\
M_{ji}^3(\mathbf{q}) &=& \frac{1}{N}\sum_{a=A,B}U_a
\mu^{\ast}_{a\mathbf{k}_i-\mathbf{q}\downarrow}\mu_{a\mathbf{k}_{i}\uparrow}\mu_{a\mathbf{k}_j-\mathbf{q}\downarrow}\mu^{\ast}_{a\mathbf{k}_{j}\uparrow}. \label{M3}
\end{eqnarray}
Here, $\varepsilon_d(\mathbf{k})$ is the dispersion of the lower electron band, and $\mu_{a\mathbf{k}\sigma}$ ($a=A,B$) the probability amplitude of sublattice $a$ that contributes to the lower band $d_{\mathbf{k}\sigma}=\sum_{a=A,B}\mu_{a\mathbf{k}\sigma}c_{a\mathbf{k}\sigma}$ (Details see supplement material). A remarkable consequence of the projection onto the lower electron band is that the dimension of the Hilbert space of spin-1 excitations scales linearly with respect to the system size \cite{Su_PRB2018,Su_PRB2019}, which is in sharp contrast to the exponential dependence met in the usual exact diagonalization. This enables us to access a much larger system. In the following, all the bulk spectra are obtained with a numerical diagonalization of Eq. (\ref{PHP}) with $N=60\times60$.

\begin{figure}
\centering
\includegraphics[width=0.48\textwidth]{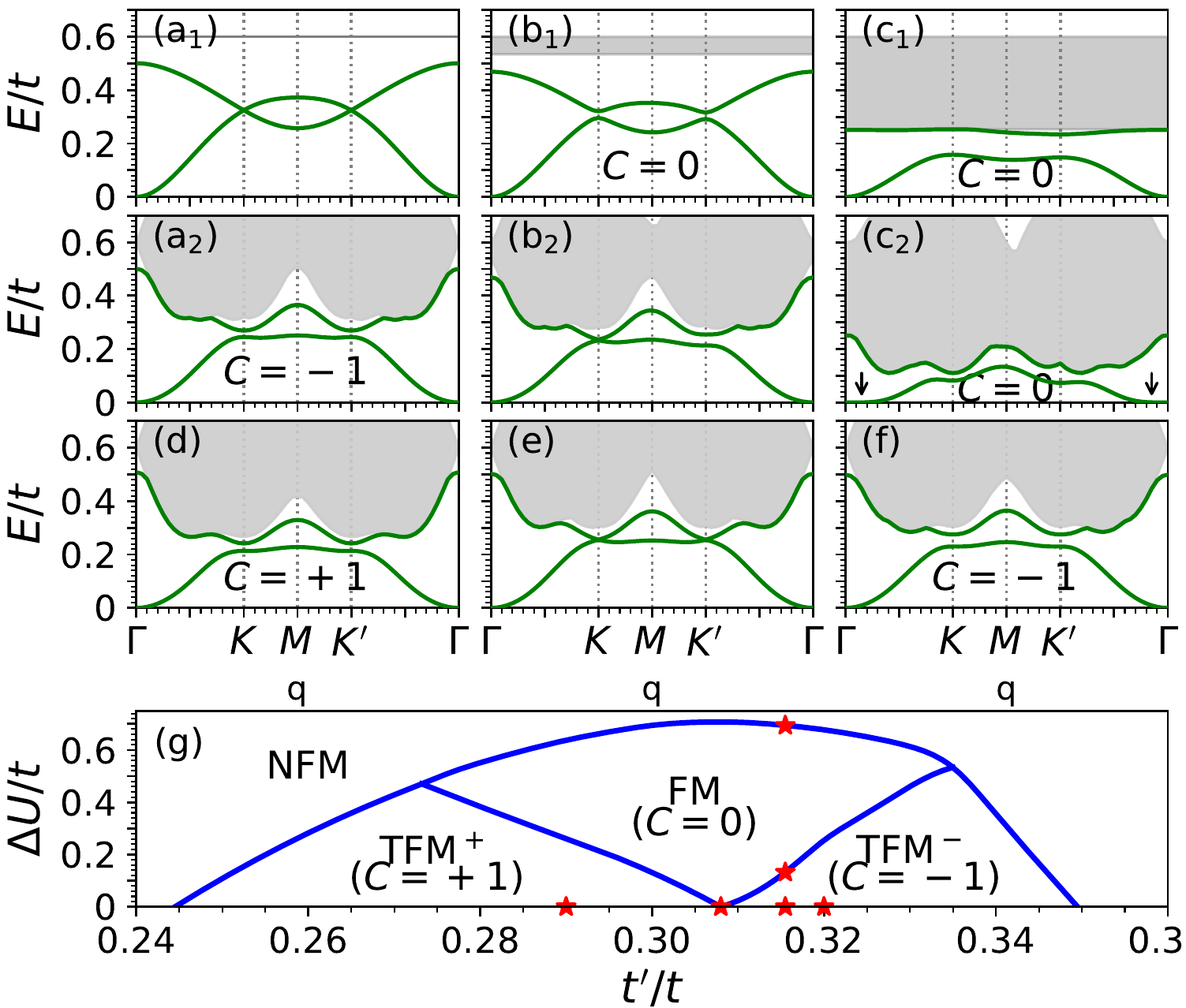}
\caption{(color online). (a)-(f): Spin excitation spectra of the $1/4$ filled Haldane-Hubbard model. (g): Its phase diagram in the $\Delta U(\equiv U_A-U_B)-t^\prime$ space. (a$_1$)-(c$_1$) are results in the flatband limit, and (a$_2$)-(f) those when the dispersion of the lower electron band is considered.
(a)-(c) are for different $U_B$ with the same $t^\prime=0.3155$: $U_B=1.2$ (a), $U_B=1.068$ (b), and $U_B=0.506$ (c). (d)-(f) are for different $t^\prime$ with the same $U_B=1.2$: $t^\prime=0.29$ (d), $t^\prime=0.308$ (e), and  $t^\prime=0.32$ (f). Other parameters are fixed at $t=1.0$, $\phi=0.656$, $U_A=1.2$. The red stars mark the parameters used in (a)-(f).}
\label{bulkresult}
\end{figure}

\par We begin with the spin excitation spectra in the flatband limit obtained by setting the $M_i^1(\mathbf{q})$ term in Eq. (\ref{PHP}) [see also Eq. (\ref{M1})] to be zero, and the results are presented in Fig. \ref{bulkresult}(a$_1$-c$_1$). The spectra consist of two parts: the low-lying spin waves labeled by the green lines and the high-energy Stoner continuum labeled by the grey region. The spin waves contain two branches of well-defined magnon bands which are attributed to be the acoustic and optical bands. When $U_A=U_B$, as shown in Fig. \ref{bulkresult}(a$_1$), the magnon bands follow quite similar dispersions to the NN tight-binding energy bands for electrons in the honeycomb lattice, in which the notable feature is the existence of Dirac points at $K$/$K^\prime$. When the imbalance of the Hubbard interactions $U_A \ne U_B$ is introduced, the Dirac magnons open gaps [Figs. \ref{bulkresult}(b$_1$)-(c$_1$)]. However, the Chern number of the acoustic branch is found to be zero, so it is still a topological trivial magnon band.

\par Then, we study the effects of the nonflatness of the lower electron band and the results are shown in Fig. \ref{bulkresult}(a$_2$-c$_2$). The nonflatness alone can also open the Dirac magnon gap [Fig. \ref{bulkresult}(a$_2$)]. Remarkably, now the Chern number of the acoustic magnon band is $-1$, so the magnon band is topologically nontrivial. From Fig. \ref{bulkresult}(b$_2$)-(c$_2$), one can see that the imbalance of the Hubbard interactions can close this topological gap and reopen a new one. But, the Chern number changes from $-1$ to 0 after the gap reopening. Another interesting observation is that the Chern number of the acoustic magnon band can be altered from $+1$ to $-1$ with the closing and reopening of the gap by tuning the NNN electron hopping, as shown in Figs. \ref{bulkresult}(d)-(f), indicating that the system changes from one topological state to another. We note that the topology of the electron bands keeps unchanged during this process. This result suggests that the topology of the magnon band does not bear a simple direct relation with that of the electron band. In addition, we find that the Chern number of the magnon band changes its sign if reverting the NNN hopping phase $\phi$. When the lower electron band acquires a dispersion, the ferromagnetic ground state possesses a tendency toward instability \cite{Tasaki_PRL1994,Su_PRB2019}, so we need to check its stability. As discussed in Ref. \cite{Su_PRB2019}, the instability can be determined by the softening of the magnon, which is signaled by the appearance of the zero value in the magnon dispersion at a finite $q$-point. From Fig. \ref{bulkresult}(c$_2$), we can identify that the acoustic branch is approaching to be zero at the $q$-point indicated by the black arrows, suggesting that the ferromagnetic state approaches the critical instability point in this case. Thus, we summarize the results by the phase diagram shown in Fig. \ref{bulkresult}(g). In the $\Delta U-t^\prime$ parameter space, where $\Delta U\equiv U_A-U_B$, we find one nonferromagnetic (NFM) phase and three ferromagnetic phases with different magnon band topologies, i.e. the trivial ferromagnetic phase (FM) with zero Chern number, and the two topological ferromagnetic phases (TFM$^+$ and TFM$^-$) with $\pm1$ Chern number, respectively.

\begin{figure}
\centering
\includegraphics[width=0.48\textwidth]{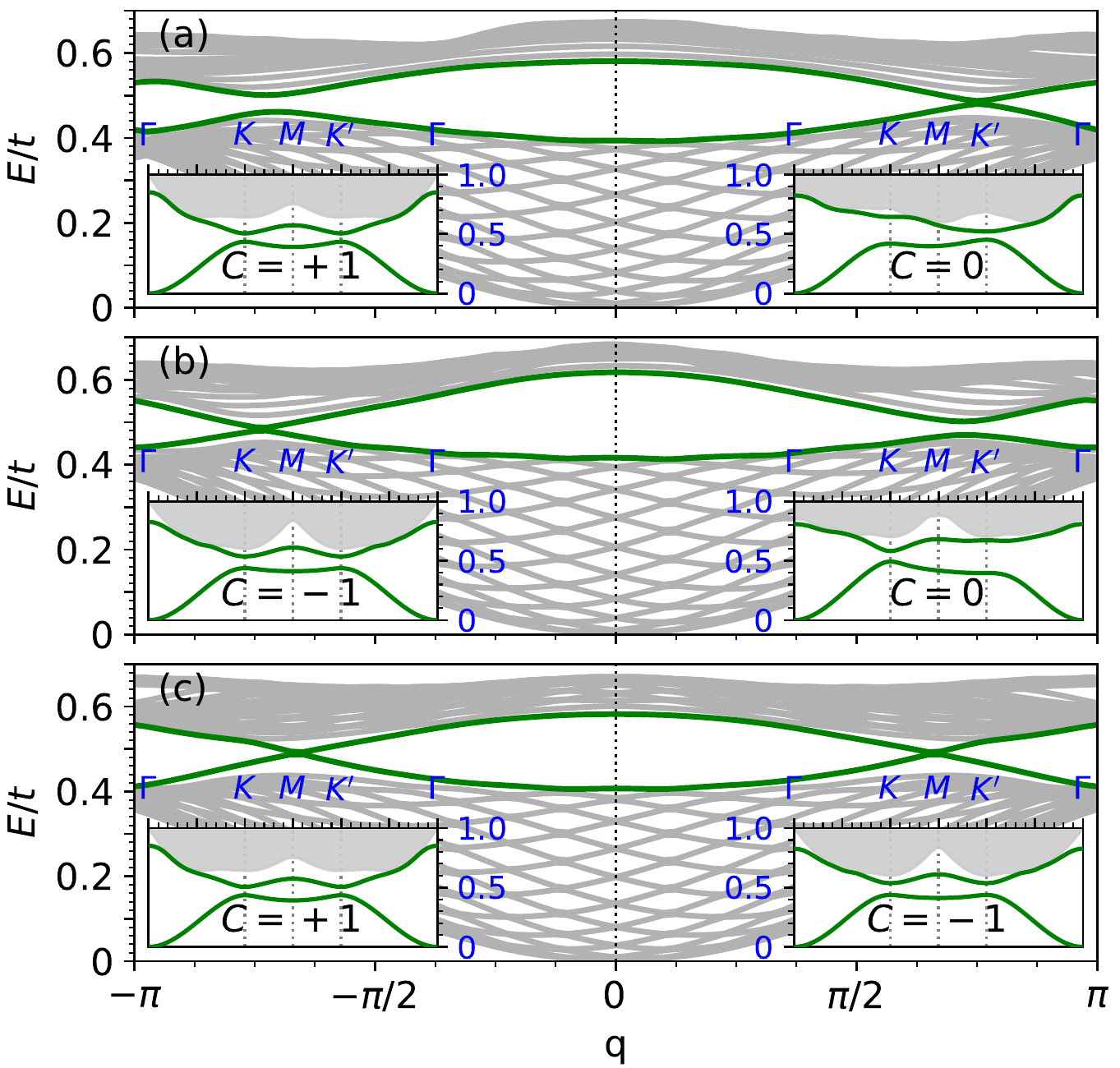}
\caption{(color online). Spin-1 excitation spectra of the $1/4$ filled Haldane-Hubbard model on magnetic domain walls with (a) $t^{\prime L}=t^{\prime R}=0.28$, $U^L_A=2.0$, $U^L_B=2.0$, $U^R_A=2.7$, $U^R_B=1.7$, (b) $t^{\prime L}=t^{\prime R}=0.33$, $U^L_A=2.0$, $U^L_B=2.0$, $U^R_A=2.7$, $U^R_B=1.7$, (c) $t^{\prime L}=0.28$, $t^{\prime R}=0.33$, $U^L_A=U^L_B=U^R_A=U^R_B=2.0$. Other parameters are fixed at $t=1.0$, $
\phi=0.656$. Insets show the corresponding bulk spin-1 excitation spectra of the left and right halves of the domain wall system.}
\label{domainwallspectra}
\end{figure}

\par According to the bulk-edge correspondence, a direct consequence of the topological magnon band is the simultaneous existence of the localized in-gap modes in the case of a open boundary condition. In the calculations to check this consequence, a difficulty arises from the electronic edge states resulting from the topological nontrivial electron band, which will cross the gap between the upper and lower electron band so that the spin fully polarized state is no more energetically favorable. As an alternative, we explore the in-gap magnon modes on magnetic domain walls, as illustrated in Fig. \ref{lattice}(c). In such a geometry, the system still assumes periodic boundary conditions on both directions, whereas along one direction (say the $x$ direction) it is composed of two halves having the same electron band topologies but different magnon band topologies by taking on different parameters. Another difficulty is due to the numerics. Because of the lack of momentum conservation along the $x$ direction, the number of unit cells along this direction we can handle is reduced to about a dozen, on the other hand, the magnon band gap is quite small (Figs. \ref{bulkresult}(a)-(f)). Thus, it is impossible to probe clearly the in-gap states. To resolve this, we artificially increase the magnon band gap by choosing much larger $U_A$ and $U_B$ \cite{gap}. Fig. \ref{domainwallspectra}(a)-(c) show the magnon spectra on TFM$^+$-FM, TFM$^-$-FM and TFM$^+$-TFM$^-$ domain walls, and the insets the corresponding bulk spin excitation spectra. One can see clearly the existence of the chiral in-gap magnon modes in all three cases, in particular, the number of the in-gap modes equals the difference of the Chern number between the two halves.

\begin{figure}
\centering
\includegraphics[width=0.48\textwidth]{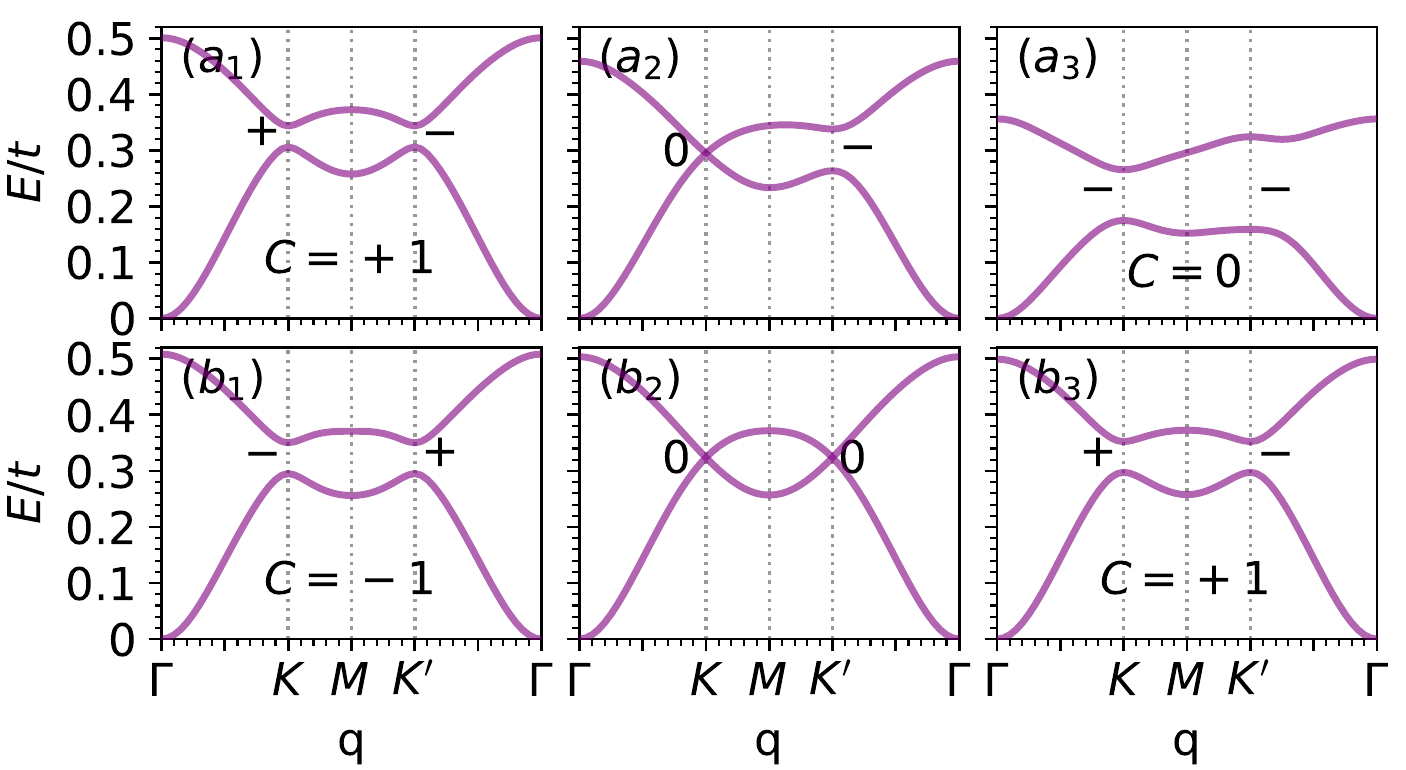}
\caption{(color online). Dispersion relations of spin waves obtained by first-order perturbation theory applied to the effective model with (a$_1$) $t^\prime=0.3155$, $U_B=1.2$, (a$_2$) $t^\prime=0.3155$, $U_B=1.0$, (a$_3$) $t^\prime=0.3155$, $U_B=0.506$, (b$_1$) $t^\prime=0.29$, $U_B=1.2$, (b$_2$) $t^\prime=0.3055$, $U_B=1.2$, (b$_3$) $t^\prime=0.32$, $U_B=1.2$. Other parameters are fixed at $t=1.0$, $\phi=0.656$, $U_A=1.2$. The $+$/$-$ marks denote the signs of the mass terms at $K$($K^\prime$) points and $0$ indicate that the mass term is zero.}
\label{effectivespectra}
\end{figure}

\par For a local spin model, a clear picture of the magnon band topology can be obtained using the LSWT, by which one can represent approximately the spin waves (magnon) by the bosonic operators and get the free bosonic model. However, the itinerant spin waves are held back by the lack of such an effective theory up to now \cite{Su_PRB2018}. From an insightful observation, we find that the $M^3_{ji}(\mathbf{q})$ term in the projected Hamiltonian Eq. (\ref{PHP}) can be decomposed as the sum of the direct product between the vectors $|v^a_i(\mathbf{q})\rangle\equiv\sqrt{\frac{U_a}{N}}\mu^{\ast}_{a\mathbf{k}_i-\mathbf{q}\downarrow}\mu_{a\mathbf{k}_{i}\uparrow}$: $M^3_{ji}(\mathbf{q})=-\sum_{a=A,B}|v^a_i(\mathbf{q})\rangle\langle v^a_j(\mathbf{q})|$. Each component of the vector $|v^a(\mathbf{q})\rangle$ is the probability amplitude to create a spin-1 particle-hole excitation in the lower electron band in the sublattice $a$, thus, we call this vector the ``sublattice particle-hole vector''. In the flatband limit, $M^2_i(\mathbf{q})=U/2$ when $U_A=U_B=U$ (see supplement material), so Eq. (\ref{PHP}) can be rewritten as a $2\times2$ matrix,
\begin{equation}\label{effective}
M^\text{Flat}_{ab}(\mathbf{q})=\frac{U}{2}\delta_{ab}-\sum_i\langle v_i^a(\mathbf{q})|v_i^b(\mathbf{q})\rangle.
\end{equation}
Thus, we find the exact bases $|v^{A,B}(\mathbf{q})\rangle$ for spin waves (magnons) in the flatband limit, on which the projected Hamiltonian is reduced to that describing the free magnons. It can be shown that $M^\text{Flat}(\mathbf{q})$ behaves as massless-Dirac-like Hamiltonians with opposite chiralities between the $K$ and $K^\prime$ points (see supplement material). When the dispersion of the electron band and the imbalance of the Hubbard interactions are considered, the terms $M^1_i(\mathbf{q})$ and $M^2_i(\mathbf{q})$ are no long zero or constants, and can be treated as perturbations to Eq. (\ref{effective}). In this case, we obtain the $2\times 2$ effective Hamiltonian up to the first order approximation. It turns out that these terms act exactly as the mass term of the Dirac magnons at $K/K^\prime$ points. In Figs. \ref{effectivespectra}(a)-(b), we plot the spectra of magnons calculated based on the effective Hamiltonian with the same parameters with those shown in Figs. \ref{bulkresult}(a$_2$)-(c$_2$) and Figs. \ref{bulkresult}(d)-(f), except in Figs. \ref{effectivespectra}(a$_2$) and (b$_2$) a slight difference in $U_B$ or $t^{\prime}$ is adopted to lead to the case of a vanishing gap. One can see that the results in Fig. \ref{effectivespectra} share similar spectra as shown in Fig. \ref{bulkresult} (The results in the flat-band limit are the same), indicating that the above perturbation treatment works here. Then, we calculate the mass term around the $K/K^\prime$ points and indicate its sign by $+$/$-$ in Fig. \ref{effectivespectra}. It shows that all topological magnon bands hosting nonzero Chern numbers have the opposite-signed mass term between the $K$ and $K^\prime$ points, while the trivial ones have the same-signed term (Details are in the supplement material). Therefore, we conclude that it is the ``mass inversion mechanism'' leading to the nontrivial magnon band topology.

\par In summary, we report the first theoretical realization of two dimensional itinerant topological magnons by numerical and analytical investigations on the quarter- filled Haldane Hubbard model with a nearly flat electron band. We find Dirac magnons in the flatband limit. Although the imbalance in the Hubbard interactions opens trivial gaps for the Dirac magnons, the magnon gap induced by the nonflatness of the electron band is topological. Correspondingly, the in-gap magnon modes are shown to exist on magnetic domain walls. We find the exact set of bases for spin waves in the flatband limit leading us to construct an effective model to explore the origin of the nontrivial magnon band which is attribute to the ``mass inversion mechanism''.

\begin{acknowledgments}
\par This work was supported by the National Natural Science Foundation of China (11774152 and 11674158) and National Key Projects for Research and Development of China (Grant No. 2016YFA0300401).
\end{acknowledgments}

\bibliography{reference}

\begin{thebibliography}{54}
\bibitem [{gap()}]{gap}%
  \BibitemOpen
  \href@noop {} {\bibinfo  {journal} {Physically, the projection onto the lower
  electron band does not apply in this case, but here it is just for an
  illustration of the numerical check for the existence of in-gap magnon
  states}\ }\BibitemShut {NoStop}%
\end{thebibliography}

\widetext
\newpage
\appendix
\section{Supplemental Material}

\setcounter{equation}{0}
\setcounter{figure}{0}
\setcounter{table}{0}
\setcounter{section}{0}
\renewcommand{\theequation}{S\arabic{equation}}
\renewcommand{\thesection}{S\arabic{section}}
\renewcommand{\thetable}{S\arabic{table}}
\renewcommand{\thefigure}{S\arabic{figure}}

\section{Haldane model and mass inversion mechanism}
\par The free part $\hat{H}_0$ of the Hamiltonian of the Haldane Hubbard model in the momentum space can be written as $\hat{H}_0=\sum_{\mathbf{k}\sigma}\psi^{\dagger}_{\mathbf{k}\sigma}H_0(\mathbf{k})\psi_{\mathbf{k}\sigma}$, where $\psi^{\dagger}_{\mathbf{k}\sigma}=(c^{\dagger}_{A\mathbf{k}\sigma},c^{\dagger}_{B\mathbf{k}\sigma})$, and
\begin{equation}\label{H0k}
H_0(\mathbf{k})=\sum_{\alpha=0,x,y,z}h_\alpha(\mathbf{k})\tau^\alpha,
\end{equation}
with $\tau^\alpha$ ($\alpha=0,x,y,z$) being the identity matrix and the three Pauli matrices in the sublattice space, and
\begin{eqnarray*}
h_0(\mathbf{k})&=&2t^\prime\cos\phi\sum_{i=1}^3\cos(\mathbf{k}\cdot\mathbf{b_i}), \\
h_x(\mathbf{k})&=&t\sum_{i=1}^3\cos(\mathbf{k}\cdot\mathbf{a}_i), \\
h_y(\mathbf{k})&=&-t\sum_{i=1}^3\sin(\mathbf{k}\cdot\mathbf{a}_i), \\
h_z(\mathbf{k})&=&-2t^\prime\sin\phi\sum_{i=1}^3\sin(\mathbf{k}\cdot\mathbf{b}_i).
\end{eqnarray*}
Here, $\mathbf{a}_1$, $\mathbf{a}_2$ and $\mathbf{a}_3$ are the displacements from a A site to its three NN B sites, defined so that $\mathbf{a}_1\times\mathbf{a}_2$ points to the positive direction of z axis, and $\mathbf{b}_1=\mathbf{a}_2-\mathbf{a}_3$, $\mathbf{b}_2=\mathbf{a}_3-\mathbf{a}_1$, $\mathbf{b}_3=\mathbf{a}_1-\mathbf{a}_2$. Note that due to the SU(2) spin rotation symmetry, $H_0(\mathbf{k})$ is independent of $\sigma$. Therefore, we will omit the spin index in the following in this section. The dispersion relation of the free electron bands reads
\begin{eqnarray*}
E^+(\mathbf{k})&=& h_0(\mathbf{k})+h(\mathbf{k}), \\
E^-(\mathbf{k})&=& h_0(\mathbf{k})-h(\mathbf{k}),
\end{eqnarray*}
where $h(\mathbf{k})=\sqrt{h^2_x(\mathbf{k})+h^2_y(\mathbf{k})+h^2_z(\mathbf{k})}$. The corresponding eigenvectors are
\begin{eqnarray}
\Psi_{E^+}(\mathbf{k})&=& \frac{1}{\sqrt{2h(\mathbf{k})\left[(h(\mathbf{k})+h_z(\mathbf{k})\right]}}
\begin{bmatrix}
h_z(\mathbf{k})+h(\mathbf{k}) \\
h_x(\mathbf{k})+ih_2(\mathbf{k})
\end{bmatrix},
\label{PsaiP}\\
\Psi_{E^-}(\mathbf{k})&=& \frac{1}{\sqrt{2h(\mathbf{k})\left[(h(\mathbf{k})-h_z(\mathbf{k})\right]}}
\begin{bmatrix}
h_z(\mathbf{k})-h(\mathbf{k})\\
h_x(\mathbf{k})+ih_2(\mathbf{k})
\end{bmatrix}\label{PsaiM}.
\end{eqnarray}
The Berry connection $A_i(\mathbf{k})$($i=k_x,k_y$) of the lower electron band is
\begin{equation*}
A_i(\mathbf{k}) = i\langle\Psi_{E^-}(\mathbf{k})|\partial_{k_i}|\Psi_{E^-}(\mathbf{k})\rangle
                         = \frac{-1}{2h\left(h-h_z\right)}\left(h_y\partial_{k_i}h_x-h_x\partial_{k_i}h_y\right),
\end{equation*}
and the corresponding Berry curvature $F(\mathbf{k})$ is
\begin{equation*}
F(\mathbf{k}) = \partial_{k_x}A_{k_x}-\partial_{k_y}A_{k_y}
              = \frac{1}{2h^3}\epsilon_{abc}h_a\partial_{k_x}h_b\partial_{k_y}h_c,
\end{equation*}
where $\epsilon_{abc}$ is the total antisymmetric tensor. When $\phi=0$ and hence $h_z(\mathbf{k})=0$, there are two inequivalent massless Dirac points of the free electron bands at $K/K^\prime=\frac{4\pi}{3}(\pm1,0)$ in the BZ. Expand Eq. (\ref{H0k}) around $K/K^\prime$ points, i.e. let $\mathbf{k}\rightarrow K/K^\prime+\mathbf{k}$ where the latter $\mathbf{k}$ is small, we find
\begin{eqnarray}
H_0^K(\mathbf{k})          &\simeq& m_0\tau^0+v_F\left(k_x\tau^x+k_y\tau^y\right)-m_z\tau^z, \label{HK1} \\
H_0^{K^\prime}(\mathbf{k}) &\simeq& m_0\tau^0+v_F\left(-k_x\tau^x+k_y\tau^y\right)+m_z\tau^z. \label{HK2}
\end{eqnarray}
Here, $m_0=-3t^\prime\cos\phi$, $v_F=-\frac{\sqrt{3}}{2}t$, and $m_z=3t^\prime\sin\phi$. Clearly, when $t^\prime\ne0$ and $\phi\ne0$, Eqs. (\ref{HK1})-(\ref{HK2}) are massive Dirac Hamiltonians with opposite-signed mass terms (the $\pm m_z\tau^z$ term). Due to the different signs of the $\tau^x$ term, these Dirac Hamiltonians have opposite chiralities as well. Correspondingly, the Berry curvature of the lower electron band around $K/K^\prime$ points becomes
\begin{equation*}
F^K(\mathbf{k})\simeq F^{K^\prime}(\mathbf{k})\simeq-\frac{v_F^2m_z}{2\left[m_z^2+(v_Fk_x)^2+(v_Fk_y)^2\right]^{3/2}}
\end{equation*}
The Chern number $C$ of the lower electron band is the sum of individual contributions $C^{K/K^\prime}$ from the above massive Dirac points:
\begin{equation*}
C^{K/K^\prime}=\frac{1}{2\pi}\int d^2\mathbf{k} F^{K/K^\prime}(\mathbf{k})=-\frac{\text{sign}(m_z)}{2}.
\end{equation*}
Therefore, $C=C^K+C^{K^\prime}=-\text{sign}(m_z)$. When $m_z\ne0$, the Chern number of the lower electron band is $\pm1$ depending on the sign of $m_z$. Following the above derivations, we want to remark that the Chern number of a single massive Dirac point is always $\pm1/2$, whose sign is determined by its chirality as well as the sign of its mass term. With an reversion of its chirality or its mass sign, the Chern number is also reverted. Thus, for a lattice model that hosts two Dirac points with opposite chiralities, a nonzero Chern number of the energy band exists if and only if the signs of the mass term at these two Dirac points are different. This is the so-called ``mass inversion mechanism''.

\section{Details on the effective model describing the spin wave excitations in 1/4 filled Haldane-Hubbard model with a nearly flat electron band}
\par When the lower electron band of the Haldane-Hubbard model is nearly flat by a proper tuning of $t^\prime$ and $\phi$, the low-energy physics is dominated by the degrees of freedom in this band if it is fractionally filled for the parameter space where the Hubbard interactions are smaller than the electron gap between the two free electron bands. Then the Hamiltonian can be projected onto this band. According to Eq. (\ref{PsaiM}), the eigen operator of this band is $d_{\mathbf{k}\sigma}=\mu_{A\mathbf{k}\sigma}c_{A\mathbf{k}\sigma}+\mu_{B\mathbf{k}\sigma}c_{B\mathbf{k}\sigma}$, with
\begin{equation*}
\mu_{A\mathbf{k}\sigma} =  \frac{h_z(\mathbf{k})-h(\mathbf{k})}{\sqrt{2h(\mathbf{k})\left[(h(\mathbf{k})-h_z(\mathbf{k})\right]}}, \;
\mu_{B\mathbf{k}\sigma} =  \frac{h_x(\mathbf{k})+ih_2(\mathbf{k})}{\sqrt{2h(\mathbf{k})\left[(h(\mathbf{k})-h_z(\mathbf{k})\right]}}.
\end{equation*}
For Eq. (\ref{M2}) in the main text, if $U_A=U_B=U$, we have
\begin{equation*}
M^2_i(\mathbf{q})=\frac{U}{N}\sum_{\mathbf{p}}\left|\mu_{A\mathbf{p}\uparrow}\right|^2\left|\mu_{A\mathbf{k}_i-\mathbf{p}\downarrow}\right|^2
                 +\left|\mu_{B\mathbf{p}\uparrow}\right|^2\left|\mu_{B\mathbf{k}_i-\mathbf{p}\downarrow}\right|^2.
\end{equation*}
Note that
\begin{equation*}
\left|\mu_{A\mathbf{p}\sigma}\right|^2 = \frac{1}{2}\left[1-\frac{h_z(\mathbf{p})}{2h(\mathbf{p})}\right],\;
\left|\mu_{B\mathbf{p}\sigma}\right|^2 = \frac{1}{2}\left[1+\frac{h_z(\mathbf{p})}{2h(\mathbf{p})}\right].
\end{equation*}
However, $h_z(-\mathbf{p})=-h_z(\mathbf{p})$ and $h(-\mathbf{p})=h(\mathbf{p})$, thus, $\sum_{\mathbf{p}}h_z(\mathbf{p})/h(\mathbf{p})=0$. Therefore,
\begin{equation*}
M^2_i(\mathbf{q})=\frac{U}{N}\sum_{\mathbf{p}}\frac{1}{2}\left|\mu_{A\mathbf{k}_i-\mathbf{p}\downarrow}\right|^2
                 +\frac{1}{2}\left|\mu_{B\mathbf{k}_i-\mathbf{p}\downarrow}\right|^2
                 =\frac{U}{2}.
\end{equation*}
On the other hand, in terms of the ``sublattice particle-hole vectors'' $|v^a_i(\mathbf{q})\rangle\equiv\sqrt{\frac{U_a}{N}}\mu^{\ast}_{a\mathbf{k}_i-\mathbf{q}\downarrow}\mu_{a\mathbf{k}_{i}\uparrow}$ ($a=A,B$), Eq. (\ref{M3}) in the main text can be written as $M^3_{ji}(\mathbf{q})=-\sum_{a=A,B}|v^a_i(\mathbf{q})\rangle\langle v^a_j(\mathbf{q})|$. Clearly, the eigenvectors with nonzero eigenvalues of $M^3(\mathbf{q})$ must be a superposition of the vectors $|v^A(\mathbf{q})\rangle$ and $|v^B(\mathbf{q})\rangle$. In the flatband limit with symmetric sublattice Hubbard interactions, Eq. (\ref{PHP}) in the main text reduces to
\begin{equation*}
M^\text{Flat}_{ji}(\mathbf{q})\equiv\left.\left[M^2(\mathbf{q})\delta_{\mathbf{k}_j,\mathbf{k}_i}+M^3_{ji}(\mathbf{q})\right]\right|_{U_A=U_B=U}
=\frac{U}{2}\delta_{\mathbf{k}_j,\mathbf{k}_i}-\sum_{a=A,B}|v^a_i(\mathbf{q})\rangle\langle v^a_j(\mathbf{q})|
\end{equation*}
Therefore, the eigenvectors of $M^\text{Flat}$ superposed by $|v^A(\mathbf{q})\rangle$ and $|v^B(\mathbf{q})\rangle$ correspond to the spin wave excitations and the eigenvectors of $M^\text{Flat}$ with $U/2$ eigenvalues correspond to the Stoner continuum. Thus, the space spanned by $|v^A(\mathbf{q})\rangle$ and $|v^B(\mathbf{q})\rangle$ defines the effective space of the spin wave excitations of $M^\text{Flat}$. In this space, $M^\text{Flat}$ reduces to a $2\times2$ matrix, which is shown by the Eq. (\ref{effective}) in the main text. When the dispersion of the lower electron band or sublattice Hubbard imbalance are considered, these terms can be treated as perturbations to Eq. (\ref{effective}). However, $|v^A(\mathbf{q})\rangle$ and $|v^B(\mathbf{q})\rangle$ are generally not orthonormal, making the evaluation of the matrix elements quite complicated. For simplicity, a set of orthonormalized bases will be used instead, which can be obtained by the standard Gram-Schmidt orthogonalization in linear algebra:
\begin{eqnarray*}
|\hat{v}^1(\mathbf{q})\rangle&=&S_{11}(\mathbf{q})|v^A(\mathbf{q})\rangle+S_{21}(\mathbf{q})|v^B(\mathbf{q})\rangle,\\
|\hat{v}^2(\mathbf{q})\rangle&=&S_{12}(\mathbf{q})|v^A(\mathbf{q})\rangle+S_{22}(\mathbf{q})|v^B(\mathbf{q})\rangle,
\end{eqnarray*}
with
\begin{equation}\label{Schmidt}
S_{11}(\mathbf{q})=\frac{1}{s_1(\mathbf{q})},\;
S_{21}(\mathbf{q})=0,\;
S_{12}(\mathbf{q})=-\frac{s_2(\mathbf{q})}{s_3(\mathbf{q})},\;
S_{22}(\mathbf{q})=\frac{1}{s_3(\mathbf{q})}.
\end{equation}
where $s_1(\mathbf{q})=\text{Norm}(|v^A(\mathbf{q})\rangle)$, $s_2(\mathbf{q})=\langle v^A(\mathbf{q})|v^B(\mathbf{q})\rangle/s_1^2$, and $s_3(\mathbf{q})=\text{Norm}(|v^B(\mathbf{q})\rangle-s_2|v^A(\mathbf{q})\rangle)$. Here $\text{Norm}(|v\rangle)$ is the function to get the norm of a vector $|v\rangle$: $\text{Norm}(|v\rangle)=\sqrt{\langle v|v\rangle}$. By use of the $\{|\hat{v}^1(\mathbf{q})\rangle,|\hat{v}^2(\mathbf{q})\rangle\}$ bases, the spin wave excitations of Eq. (\ref{PHP}) up to the first order perturbation approximation can be described by the following matrix:
\begin{equation}\label{Effective1st}
M^{\text{1st}}_{\alpha\beta}(\mathbf{q})=M^0_{\alpha\beta}(\mathbf{q})+M^\Omega_{\alpha\beta}(\mathbf{q})+M^{dU}_{\alpha\beta}(\mathbf{q}),
\end{equation}
with
\begin{eqnarray}
M^0_{\alpha\beta}(\mathbf{q}) &=& \frac{U_A+U_B}{2}\delta_{\alpha\beta}+\left(\left[S^\dagger(\mathbf{q}) S(\mathbf{q})\right]^{-1}\right)_{\alpha\beta}, \\
M^\Omega_{\alpha\beta}(\mathbf{q}) &=& \sum_i\langle\hat{v}^\alpha_i(\mathbf{q})
                                     |\epsilon_d(\mathbf{k}_i-\mathbf{q})-\epsilon(\mathbf{k}_i)|
                                     \hat{v}^\beta_i(\mathbf{q})\rangle, \\
M^{dU}_{\alpha\beta}(\mathbf{q}) &=& \frac{U_A-U_B}{2N}\sum_{s=A,B}\sum_i(-1)^s\langle\hat{v}^\alpha_i(\mathbf{q})
                                 |\sum_{\mathbf{p}}\left|\mu_{s\mathbf{p}\uparrow}\mu_{s\mathbf{k}_{i}-\mathbf{q}\downarrow}\right|^2|
                                 \hat{v}^\beta_i(\mathbf{q})\rangle,
\end{eqnarray}
where the $S$ matrix is defined by Eq. (\ref{Schmidt}), and $(-1)^A\equiv1$, $(-1)^B\equiv-1$. $M^{\text{1st}}(\mathbf{q})$ is a $2\times2$ Hermitian matrix, therefore, it can be expressed as
\begin{equation}\label{M}
M^{\text{1st}}(\mathbf{q})=\sum_{\alpha=0,x,y,z}m^\alpha(\mathbf{q})\sigma^\alpha
\end{equation}
with $\sigma^\alpha$($\alpha=0,x,y,z$) the identity matrix and the three Pauli matrices in the $\{|\hat{v}^1(\mathbf{q})\rangle,|\hat{v}^2(\mathbf{q})\rangle\}$ space. In theory, with Eqs. (\ref{Schmidt})-(\ref{M}), the analytical expressions for $m^\alpha(\mathbf{q})$ can be obtained. However, these expressions are too complicated to analyze, thus, we will only compute them numerically and study the asymptotic behavior of Eq. (\ref{M}) around the $K/K^\prime$ points, which is enough for the exploration of the topological properties of the magnon bands.
\par (1) When $U_A=U_B=U>0$ and in the flatband limit, we find two massless Dirac-like effective Hamiltonians with opposite chiralities near the $K/K^\prime$ points:
\begin{eqnarray*}
M^{\text{1st}}_K(\mathbf{q})          &\simeq& v_0\sigma^0+v_xq_x\sigma^x+v_yq_y\sigma^y, \\
M^{\text{1st}}_{K^\prime}(\mathbf{q}) &\simeq& v_0\sigma^0+v_xq_x\sigma^x-v_yq_y\sigma^y.
\end{eqnarray*}
where $v_0$, $v_x$ and $v_y$ are some constants dependent on the model parameters.
\par (2) When $U_A=U_B=U>0$ and the dispersion of the lower electron band is considered, we find two massive Dirac-like effective Hamiltonians with opposite chiralities as well as opposite mass terms near the $K/K^\prime$ points:
\begin{eqnarray*}
M^{\text{1st}}_K(\mathbf{q})          &\simeq& v_0^\prime\sigma^0+v_x^\prime q_x\sigma^x+v_y^\prime q_y\sigma^y+v_z\sigma^z, \\
M^{\text{1st}}_{K^\prime}(\mathbf{q}) &\simeq& v_0^\prime\sigma^0+v_x^\prime q_x\sigma^x-v_y^\prime q_y\sigma^y-v_z\sigma^z.
\end{eqnarray*}
where $v_0^\prime$, $v_x^\prime$ and $v_y^\prime$ and $v_z$ are some constants dependent on the model parameters. Note that the fermi velocity of the Dirac magnons are also renormalized by the nonflatness of the lower electron band. In this case, according to the ``mass inversion mechanism'', the magnon band are topological with a nonzero ($\pm1$) Chern number.
\par (3) When $U_A\ne U_B>0$ and in the flatband limit, we find two massive Dirac-like effective Hamiltonians with opposite chiralities but with the same-signed mass terms near the $K/K^\prime$ points:
\begin{eqnarray*}
M^{\text{1st}}_K(\mathbf{q})          &\simeq& v_0^K\sigma^0+v_x^K q_x\sigma^x+v_y^K q_y\sigma^y+v^K_z\sigma^z, \\
M^{\text{1st}}_{K^\prime}(\mathbf{q}) &\simeq& v_0^{K^\prime}\sigma^0+v_x^{K\prime} q_x\sigma^x-v_y^{K\prime} q_y\sigma^y+v_z^{K\prime}\sigma^z.
\end{eqnarray*}
where $v_0^{K/K^\prime}$, $v_x^{K/K^\prime}$, $v_y^{K/K^\prime}$, and $v_z^{K/K^\prime}$ are some constants dependent on the model parameters. Note that at $K$ and $K^\prime$ points, these parameters bear asymmetry in the amplitude but share the same sign. According to the ``mass inversion mechanism'', the magnon band for this case are topologically trivial with zero Chern number.
\par (4) When $U_A\ne U_B>0$ and the dispersion of the lower electron band is considered, we still find two massive Dirac-like effective Hamiltonians with opposite chiralities. But now the signs of the mass terms are dependent on both $\Delta U$ and the nonflatness of the lower electron band. When $\Delta U$ is dominant, the mass terms at $K/K^\prime$ points have the same sign, and the magnon band is trivial while when the nonflatness of the lower electron band is dominant, the mass terms at $K/K^\prime$ points have opposite signs and the magnon band is topological.

\end{document}